\documentstyle[epsf]{kapproc}

\kluwerbib
\let\lcitebracket(
\let\rcitebracket)
\startauthorindex

\begin{document}

\articletitle{Binaries and Globular Cluster\\ Dynamics}
 
\author{Frederic A.\ Rasio, John M.\ Fregeau, \& Kriten J.\ Joshi}
\affil{Dept of Physics, MIT, Cambridge, MA 02139, USA}

\begin{abstract} 
We summarize the results of recent theoretical work on the dynamical 
evolution of globular clusters containing primordial binaries.
Even a very small initial binary fraction (e.g., 10\%) can play a key role in
supporting a cluster against gravothermal collapse for many
relaxation times.
Inelastic encounters between binaries and single stars or other
binaries provide a very significant energy source for the cluster. 
These dynamical interactions also lead to the production of large
numbers of exotic systems such as ultracompact X-ray binaries,
recycled radio pulsars, double degenerate systems, and blue stragglers.
Our work is based on a new parallel supercomputer code
implementing H\'enon's Monte Carlo
method for simulating the dynamical evolution of dense stellar
systems in the Fokker-Planck approximation. This new code allows us to 
calculate very accurately
the evolution of a cluster containing a realistic number of stars
($N\sim 10^5-10^6$) in typically a few hours to a few days of 
computing time. The discrete, star-by-star
representation of the cluster in the simulation makes it
possible to treat
naturally a number of important processes, including
single and binary star evolution, all dynamical interactions of single
stars and binaries, and tidal interactions with the Galaxy.
\end{abstract}

\section{Introduction}

The dynamical evolution of dense star clusters is a problem of
fundamental importance in theoretical astrophysics, but many
aspects of the problem have remained unresolved in spite of years
of numerical work and improved observational data (see Meylan \&
Heggie 1997 for a recent review).
The realization over the last 10 years that primordial binaries are
present in globular clusters in dynamically significant numbers has completely
changed our theoretical perspective on these systems (see, e.g., Gao et al.\ 1991;
Hut et al.\ 1992; Sigurdsson \& Phinney 1995).
Most importantly,
dynamical interactions between hard primordial binaries and other single
stars or binaries are now thought to be the primary mechanism for supporting
a globular cluster against gravothermal contraction and avoiding core collapse.
In addition, exchange interactions
between primordial binaries and compact objects can
explain very naturally the formation of large
numbers of X-ray binaries and recycled pulsars in globular cluster cores
(see, e.g., Camilo et al.\ 2000).
Dynamical interactions involving primordial binaries can also result in
dramatically
increased collision rates in globular clusters. This is because the
interactions
are often {\it resonant\/}, with all the stars involved remaining together in
a small volume for a long time ($\sim10^2-10^3$ orbital times).
For example, in the case of an interaction between two typical hard binaries
with semi-major axes $\sim1\,$AU containing $\sim1\,M_\odot$ main-sequence
stars, the effective cross section for a direct collision between
any two of the four stars involved is essentially equal to the entire
geometric cross section of the binaries (Bacon, Sigurdsson \& Davies 1996;
Cheung, Portegies Zwart, \& Rasio 2001).
This implies a collision rate $\sim100$ times larger than for single stars.
Direct observational evidence for stellar collisions and mergers in globular
clusters comes from the detection of large numbers of blue stragglers
concentrated in the dense cluster cores (see, e.g., Bailyn 1995).

\section{Monte Carlo Simulations of Cluster Dynamics}

The first Monte Carlo methods for calculating the dynamical evolution 
of star clusters
in the Fokker-Planck approximation were developed more than 30 years ago. They
were first used to study the development of the gravothermal instability 
(H\'enon 1971a,b; Spitzer \& Hart 1971a,b). 
More recent implementations have established the Monte Carlo method
as an important
alternative to direct $N$-body integrations (see Spitzer 1987 for an
overview).
The main motivation for our recent work at MIT was our realization a few
years ago that the  latest generation of
parallel supercomputers now make it possible
to perform Monte Carlo simulations for a number of objects equal 
to the actual number of stars in a globular cluster (in contrast,
earlier work was limited to using  a small number of representative
``superstars,'' and was often plagued by high levels of numerical noise).
Therefore, the Monte Carlo method allows us to do right now what remains
an elusive goal for $N$-body simulations (see, e.g.,  Aarseth 1999): 
perform realistic, star-by-star computer simulations of globular cluster evolution.
Using the correct number of stars in a dynamical simulation
ensures that the relative rates of different dynamical processes (which all
scale differently with the number of stars) are correct. This is
particularly crucial if many different dynamical processes are to be 
incorporated, as must be done in realistic simulations.

Our implementation of the Monte Carlo method is described in detail in
the papers by 
Joshi, Rasio, \& Portegies Zwart (2000), Joshi, Nave, \& Rasio (2001), and
Joshi, Portegies Zwart, \& Rasio (2001).
We adopt the usual assumptions of spherical symmetry (with a 2D phase space 
distribution function $f(E,J)$, i.e., we do {\it not\/} assume isotropy)
and standard two-body relaxation in the weak scattering limit (Fokker-Planck 
approximation). In its simplest version, our code computes the dynamical 
evolution of a self-gravitating spherical cluster of $N$ point masses 
whose orbits 
in the cluster are specified by an energy $E$ and angular momentum $J$, 
with perturbations $\Delta E$ and $\Delta J$ evaluated on a timestep that 
is a fraction of the local two-body relaxation time. The cluster is assumed to
remain always very close to dynamical equilibrium (i.e., the relaxation time
must remain much longer than the dynamical time). 
Our main improvements over H\'enon's original method
are the parallelization of the basic algorithm and the development of a more 
sophisticated method for determining the timesteps 
and for computing the two-body relaxation from representative encounters
between neighboring stars. Our new method allows the timesteps to be made 
much smaller in order to resolve the dynamics in the cluster core more 
accurately.

We have performed
a large number of test calculations and comparisons with direct $N$-body
integrations, as well as direct integrations of the Fokker-Planck equation in phase 
space, to establish the accuracy of our basic treatment of two-body relaxation
(Joshi et al.\ 2000). Fig.~1 shows the results from a typical comparison 
between Monte Carlo and $N$-body simulations.
Fig.~2 shows the results obtained with our code for Heggie's Collaborative
Experiment run (Heggie et al.\ 1998).

\begin{figure}
\epsfxsize 4.2truein
\epsfbox{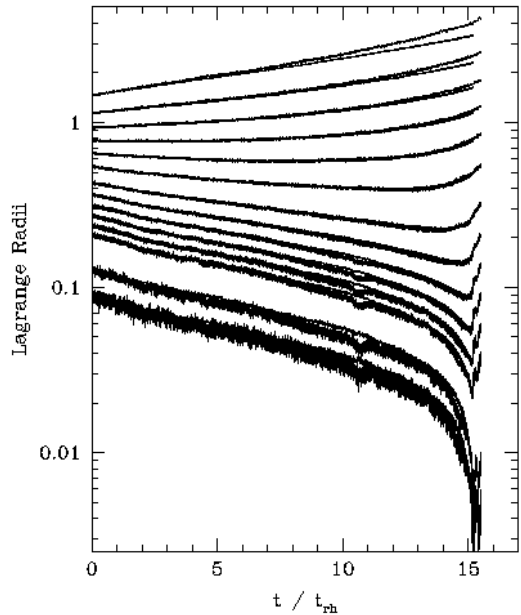}
\caption{Evolution of the Lagrange radii for an isolated, single-component
Plummer model
(from bottom to top: radii containing 0.35\%, 1\%, 3.5\%, 5\%, 7\%, 10\%, 
14\%, 20\%, 30\%, 40\%, 50\%, 60\%, 70\%, and 80\% percent of the 
total mass are shown as a function of time, 
given in units of the initial half-mass relaxation time).
The results from
a direct $N$-body integration with $N=16,384$ (noisier lines)
and from a Monte Carlo integration with $N=10^5$ stars (smoother lines)
are compared. The Monte Carlo simulation was completed in less than a day
on a Cray/SGI Origin2000 parallel supercomputer, while the $N$-body
integration ran for over a month on a dedicated GRAPE-4 computer. 
The agreement between the $N$-body and Monte Carlo results
is excellent over the entire range of Lagrange radii and time.  
The small discrepancy in the outer Lagrange radii is caused mainly
by a different treatment of escaping stars in the two models. 
In the Monte Carlo model, escaping stars are removed from the 
simulation and therefore not included in the determination of the 
Lagrange radii, whereas in the $N$-body model escaping stars are 
not removed. Note also that the Monte Carlo simulation is terminated 
at core collapse, while the $N$-body simulation continues beyond 
core collapse.}
\end{figure}

\begin{figure}
\epsfxsize 4.5truein
\epsfbox{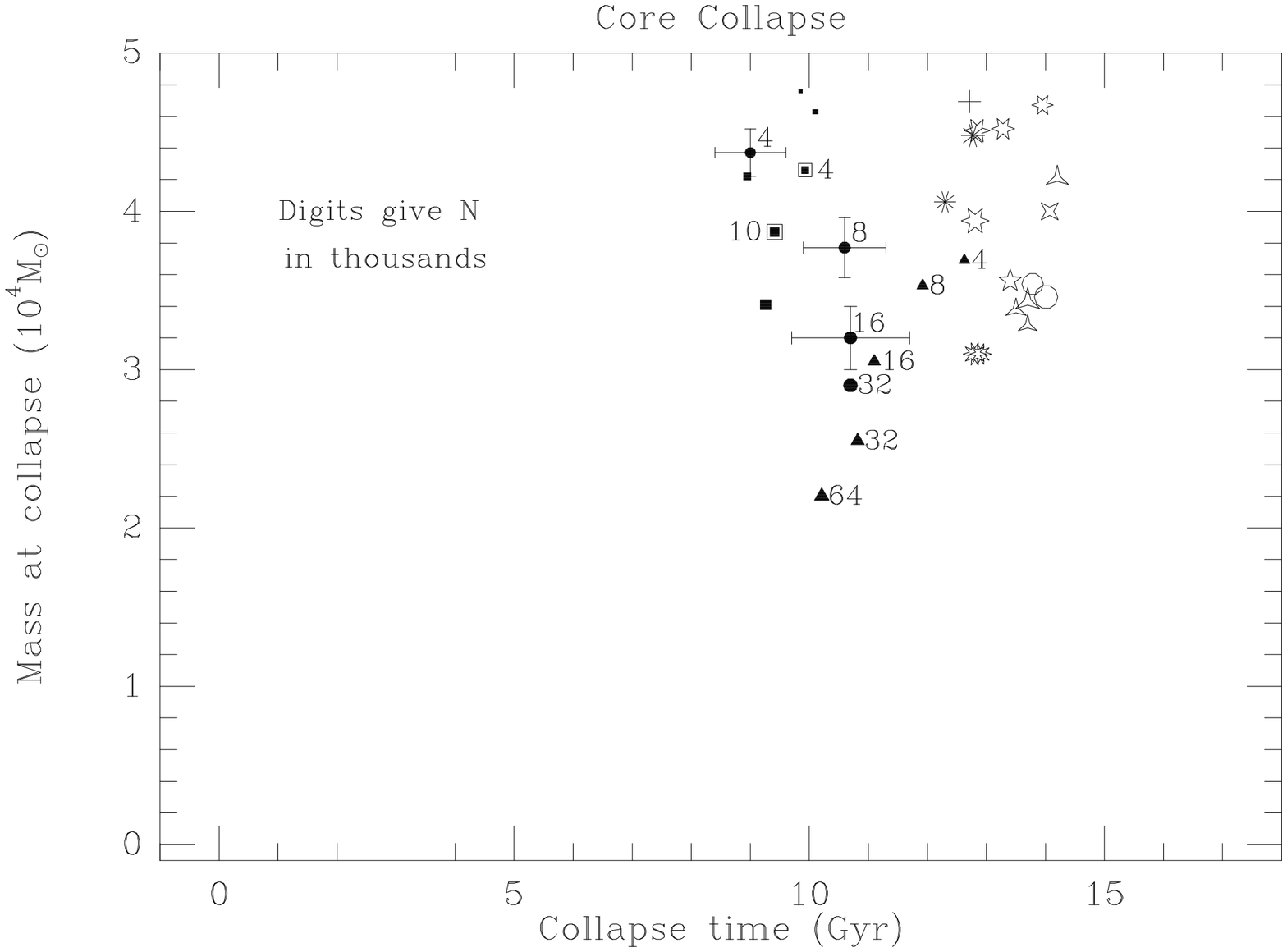}
\caption{Core collapse time $t_{\rm cc}$ and cluster mass at core collapse
$M_{\rm cc}$ as computed by many different numerical codes in Heggie's
Collaborative Experiment, and with our Monte Carlo code.
The solid round dots, triangles and squares are from various sets
of direct $N$-body simulations. Open symbols are from Fokker-Planck 
simulations and the stars are from anisotropic gas models.
Our data point is indicated by the plus symbol, corresponding to
a core collapse time $t_{\rm cc}= 12.86\,$Gyr and a mass at core
collapse $M_{\rm cc} = 4.73 \times 10^4 \,M_\odot$. The initial
condition is a King model with a dimensionless central potential
$W_0=3$, a tidal radius $r_{\rm t} = 30\,$pc, and containing a
mass $M = 6\times 10^4\,M_\odot$. The cluster contains single
stars only with a power-law IMF of slope $-2.35$ (Salpeter mass function)
between $0.1\,M_\odot$ and $1.5\,M_\odot$.
All simulations are done assuming no stellar evolution. Heating by ``3-body 
binaries'' is not included in our Monte Carlo
simulation (this only affects the evolution beyond core collapse).
See Heggie et al.\ 1998 for more details.}
\end{figure}

\section{Summary of Recent Results}

Our recent work has focused on the addition of more realistic stellar and binary
processes to the basic Monte Carlo code, as well as a simple but accurate 
implementation of a static tidal boundary in the Galactic field (Joshi et al.\
2001a). As a first application, we have studied the dependence on initial
conditions of globular cluster lifetimes in the Galactic environment.
As in previous Fokker-Planck studies (Chernoff \& Weinberg 1990;
Takahashi \& Portegies Zwart 1998),
we include the effects of a power-law initial mass function (IMF), mass loss 
through a tidal boundary, and single star evolution, and we
consider initial King models with varying central concentrations.
We find that the disruption and core-collapse times of our models are 
significantly longer than those obtained with previous 1D (isotropic) 
Fokker-Planck calculations, but agree well with more recent results
from direct $N$-body simulations and 2D Fokker-Planck integrations.
In agreement with previous studies,
our results show that the direct mass loss due to stellar evolution 
causes most clusters with a low initial central concentration 
to disrupt quickly in the Galactic tidal field. The disruption is
particularly rapid for clusters with a relatively flat IMF. 
Only clusters born with high central concentrations 
or with very steep IMFs 
are likely to survive to the present and undergo core collapse. 

In another recent study, we have used our Monte Carlo code to examine the
development of the Spitzer ``mass stratification instability'' in simple
two-component clusters (Watters, Joshi, \& Rasio 2000).
We have performed a large number of
dynamical simulations for star clusters containing two stellar populations 
with individual masses 
$m_1$ and $m_2 > m_1$, and total masses $M_1$ and $M_2 < M_1$.  We use both 
King and Plummer model initial conditions and we perform simulations for a wide
range of individual and total mass ratios, $m_2/m_1$ and $M_2/M_1$, in order to
determine the precise location of the stability boundary in this 2D parameter
space. As predicted originally by Spitzer (1969) using simple analytic arguments, we
find that unstable systems never reach energy equipartition, and are
driven to rapid core collapse by the heavier component.
These results have important implications for
the dynamical evolution of any population of primordial
black holes or neutron stars in globular clusters. In particular,
primordial black holes with $m_2/m_1\sim10$ are expected to undergo 
very rapid core collapse independent of the background cluster, and 
to be ejected from the cluster through dynamical interactions 
between single and binary black holes (see Portegies Zwart \& McMillan 2000
and references therein).
We have also used Monte Carlo simulations of simple two-component systems
to study the evaporation (or retention) of {\it low-mass\/} objects in globular clusters,
motivated by the surprising recent observations of planets and brown dwarfs in 
several clusters (Fregeau et al.\ 2001).

\begin{figure}
\epsfxsize 4.5truein
\epsfbox{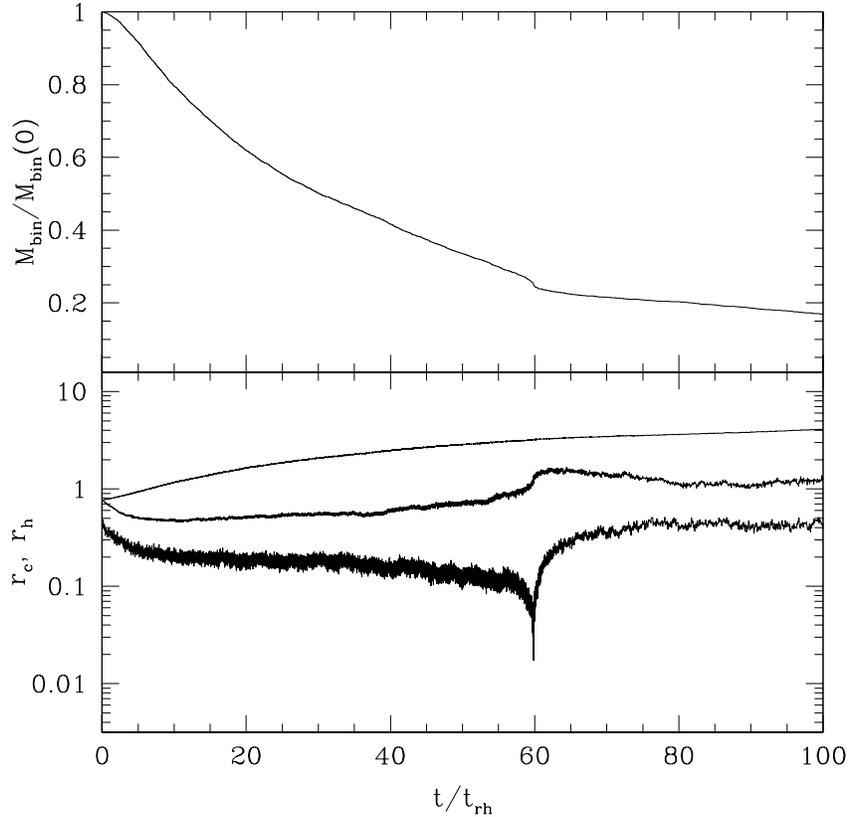}
\caption{Results of a Monte Carlo simulation for the evolution of an isolated
Plummer model containing $N=3\times10^5$ equal-mass stars, with 10\% of the stars
in primordial binaries. The binaries are initially distributed uniformly throughout 
the cluster, and with a uniform distribution in the logarithm of the binding
energy (roughly between contact and the hard-soft boundary, i.e., no soft binaries
are included). The simulation
includes a treatment of energy production and binary destruction through
binary-single and binary-binary interactions. Stellar evolution and tidal
interactions with the Galaxy are not included.
Time is given in units of the initial half-mass relaxation time $t_{\rm rh}$.
The upper panel shows the evolution of the total mass (or number) of binaries.
The lower panel shows, from top to bottom initially, the half-mass radius
of the entire cluster, the half-mass radius of the binaries, and the cluster 
core radius. These quantities are in units of the virial radius of the cluster.
Note the long, quasi-equilibrium phase of ``binary burning'' 
lasting until $t\simeq60\,t_{\rm rh}$, followed by a brief
episode of core contraction and re-expansion to an even longer quasi-equilibrium
phase with an even larger core. By $t\sim 100\,t_{\rm rh}$, only about 
15\% of the initial population of binaries remains in the cluster, but this
is enough to support the cluster against core collapse for another
$\sim 100\,t_{\rm rh}$. For most globular clusters, $t_{\rm rh}\sim10^9\,$yr, and
this is well beyond a Hubble time. The evolution shown here should be contrasted
to that of an identical cluster, but containing single stars only (Fig.~1), where
core collapse is reached at $t\simeq15\,t_{\rm rh}$.}
\end{figure}

\begin{figure}
\epsfxsize 4.5truein
\epsfbox{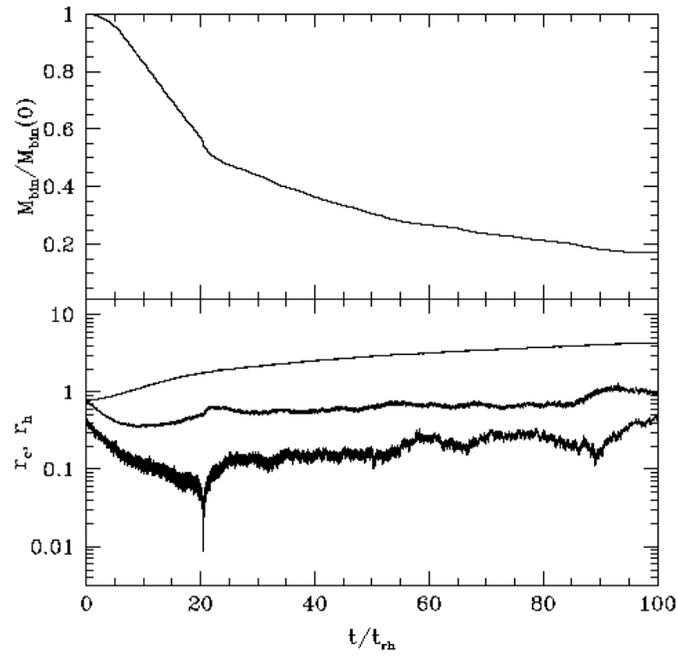}
\caption{Same as Fig.~3 but for $N=10^5$ equal-mass stars with 10\% binaries.
Note the somewhat faster and deeper initial contraction, but still followed by
re-expansion into a long-lived, quasi-equilibrium phase of binary burning.}
\end{figure}

\begin{figure}
\epsfxsize 4.5truein
\epsfbox{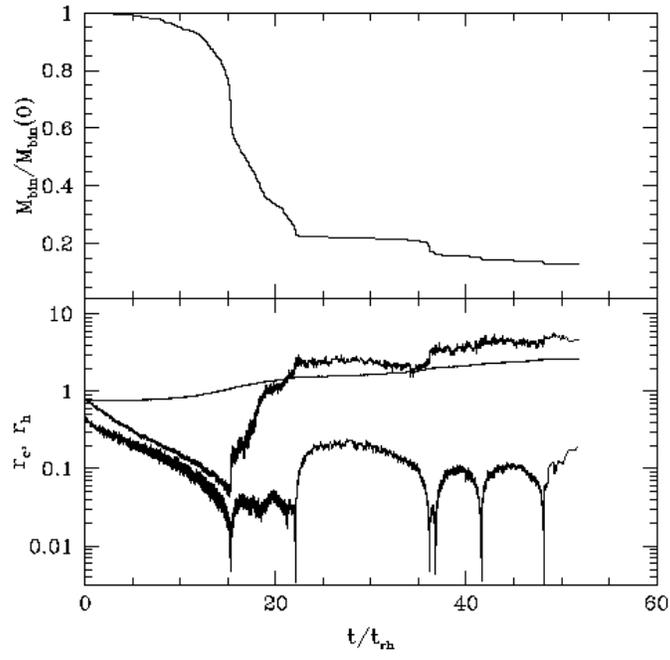}
\caption{Same as Fig.~3 but for $N=10^5$ equal-mass stars with only 1\% binaries.
Here the initial core contraction is followed by more rapid gravothermal 
oscillations, but the time-averaged core size remains similar to what is
seen in Figs.~3 and~4.}
\end{figure}

\begin{figure}
\epsfxsize 4.5truein
\epsfbox{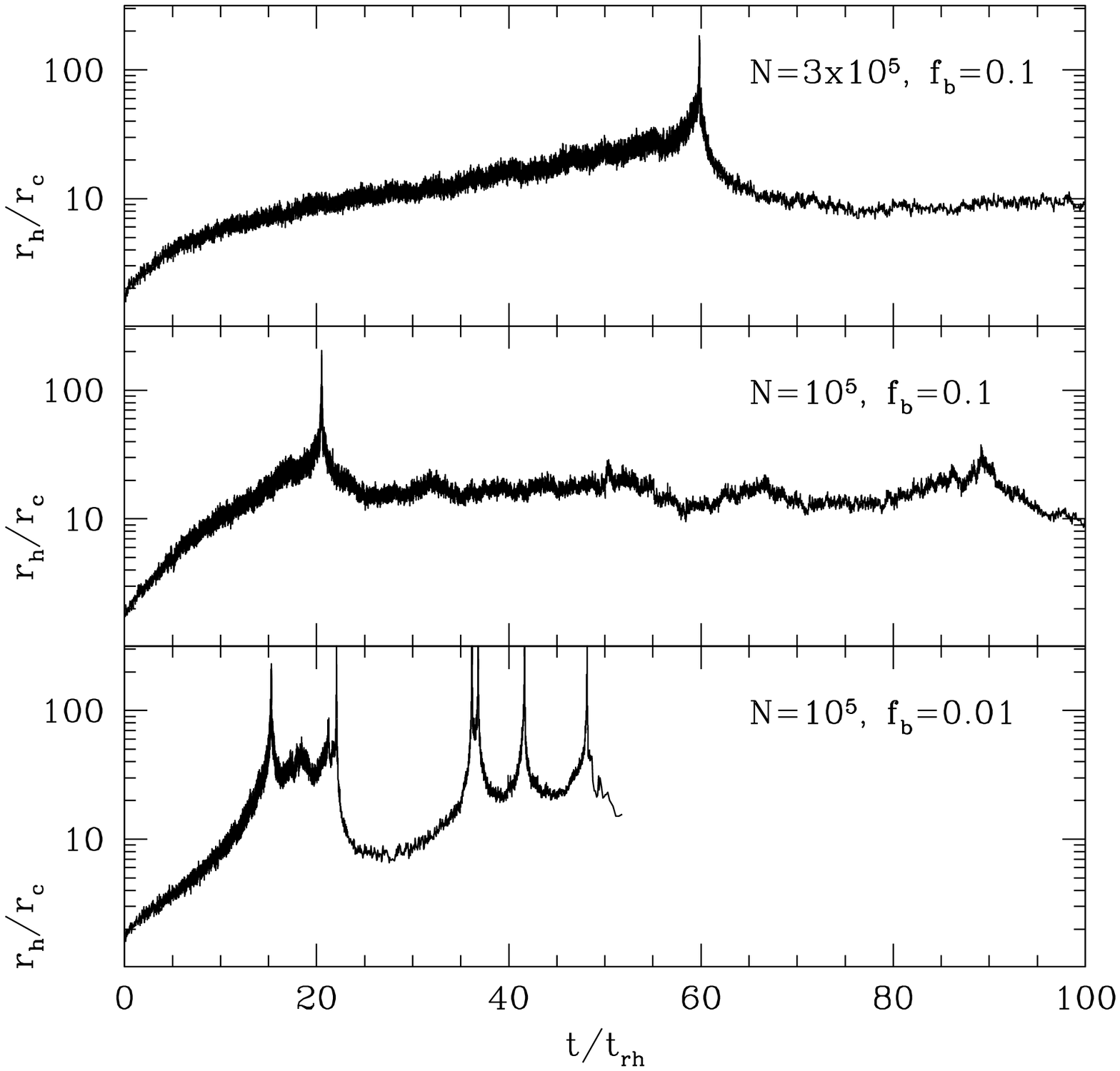}
\caption{Evolution of the half-mass radius to core radius ratio for
the three cases illustrated in Figs.~3--5. For comparison, most observed 
globular clusters with resolved cores have $r_{\rm h}/r_{\rm c}\simeq 2-10$.}
\end{figure}

Much of our current work concerns the treatment of dynamical interactions with
primordial binaries. We are in the process of completing a first study of globular
cluster evolution with primordial binaries (Joshi et al.\ 2001b), 
based on the same set of approximate
cross sections and recipes for dynamical interactions
used in the Fokker-Planck simulations of Gao et al.\ (1991).
Typical results are illustrated in Figs.~3--6. The heating of the cluster core generated
by a small population of primordial binaries can support the cluster against core
collapse for very long times, although the details of the evolution depend sensitively
both on the number of stars $N$ and on the initial binary fraction $f_{\rm b}$. 
Clusters with
smaller $N$ tend to evolve faster towards deeper core collapse (compare Figs.~3 and~4,
and note that the main dependence on $N$ through $t_{\rm rh}\propto N/\log N$ has been
scaled out), and so do clusters with smaller initial binary fractions (Fig.~5). 
After the initial core collapse, gravothermal oscillations powered by primordial
binaries can continue for very long times $\sim50-100\, t_{\rm rh}$ even with an
initial binary fraction as small as 1\%. With an initial binary fraction of
10\%, we observe a single, very moderate phase of core collapse during which the
core shrinks by a factor $\sim5-10$ in radius and then re-expands rapidly on a
timescale of a few central relaxation times (this occurs at $t/t_{\rm rh}\simeq 60$
in Fig.~3 and at $t/t_{\rm rh}\simeq 20$ in Fig.~4). The cluster then enters a second
quasi-equilibrium phase of primordial binary burning with the core radius {\it increasing
slowly\/} until well beyond $\sim100\,t_{\rm rh}$.

Since $t_{\rm rh}\sim10^9\,$yr for most clusters, if the type of evolution 
illustrated in Fig.~3 applied to all globular
clusters, there would be no ``core-collapsed'' clusters in the Galaxy
(10--15\% of all Galactic globular clusters are classified observationally as 
``core collapsed''). However, the timescale
on which real clusters will exhaust their primordial binary supply and undergo
(deeper) core collapse depends on a number of factors not considered here, including
the orbit of the cluster in the Galaxy
(the simulations are for an isolated cluster, but mass loss and
tidal shocking can accelerate the evolution dramatically), and the stellar IMF
(the clusters shown here contain all equal-mass stars and binaries;
a more realistic mass spectrum will also accelerate the evolution).
In addition, some clusters may be formed with much fewer primordial binaries
than even the 1\% considered in Fig.~5.
However, the simple picture that emerges from these simulations may well, 
to first approximation,
describe the dynamical state of most Galactic globular clusters observed today.
Indeed, for a cluster in the stable ``binary burning'' phase,
the ratio of half-mass radius to core radius $r_{\rm h}/r_{\rm c}\sim
2-10$ (Fig.~6), which is precisely 
the range of values observed for the $\sim80\%$ of globular
clusters that have a well-resolved core and are well-fitted by King models
(see, e.g., Djorgovski 1993).
Some of these clusters may have gone in the past through a brief episode of 
``moderate core
collapse'' (as shown around $t\simeq60\,t_{\rm rh}$ in Fig.~3). Yet, 
they should not be called ``core-collapsed'' (nor would they be
classified as such by observers). Unfortunately some theorists will even call
``core-collapsed'' clusters that have just reached the {\it initial phase\/} of
binary burning ($t\simeq 10-50\,t_{\rm rh}$ in Fig.~3). Since the core has 
just barely contracted by a factor $\sim2-3$ by the time it reaches this
phase, it seems hardly justified to speak of a ``collapsed'' state. 

The addition of binary stellar evolution processes to our simulations
will allow us to study in detail
the dynamical formation mechanisms for many exotic objects such as X-ray binaries
and millisecond radio pulsars, which have been detected
in large numbers in globular clusters. For example, exchange interactions
between neutron stars and primordial binaries can lead to common-envelope
systems and the formation of short-period neutron-star / white-dwarf binaries
that can become visible both as ultracompact X-ray binaries and binary millisecond
pulsars with low-mass companions (see, e.g., Camilo et al.\ 2000, on observations
of 20 such millisecond radio pulsars in 47~Tuc). Rasio, Pfahl, \& Rappaport (2000) 
present a preliminary study of this formation scenario, based on simplified
dynamical Monte Carlo simulations. In a dense cluster
such as 47~Tuc, most neutron stars acquire binary companions
through exchange interactions with primordial binaries.
The resulting systems have semimajor axes in the range $\sim 0.1-1\,$AU
and neutron star companion masses $\sim 1-3\,M_\odot$.
For many of these systems it is found that, when the companion
evolves off the main sequence and fills its Roche lobe, the subsequent
mass transfer is dynamically unstable. This leads
to a common envelope phase and the formation of short-period
neutron-star / white-dwarf binaries. For a significant fraction
of these binaries,
the decay of the orbit due to gravitational radiation will be followed
by a period of stable mass transfer
driven by a combination of gravitational radiation and tidal heating
of the companion. The properties of the resulting short-period
binaries match well those of observed binary pulsars
in 47~Tuc (Fig.~7). A similar dynamical scenario involving massive CO white
dwarfs in place of neutron stars could explain the recent detections of
double degenerate binaries containing He white dwarfs concentrated in the
core of a dense globular cluster (Hansen et al.\ 2001; see Edmonds et al.\ 1999
and Taylor et al.\ 2001 on observations of He white dwarfs in the core of
NGC 6397).

\begin{figure}
\epsfxsize 4.5truein
\epsfbox{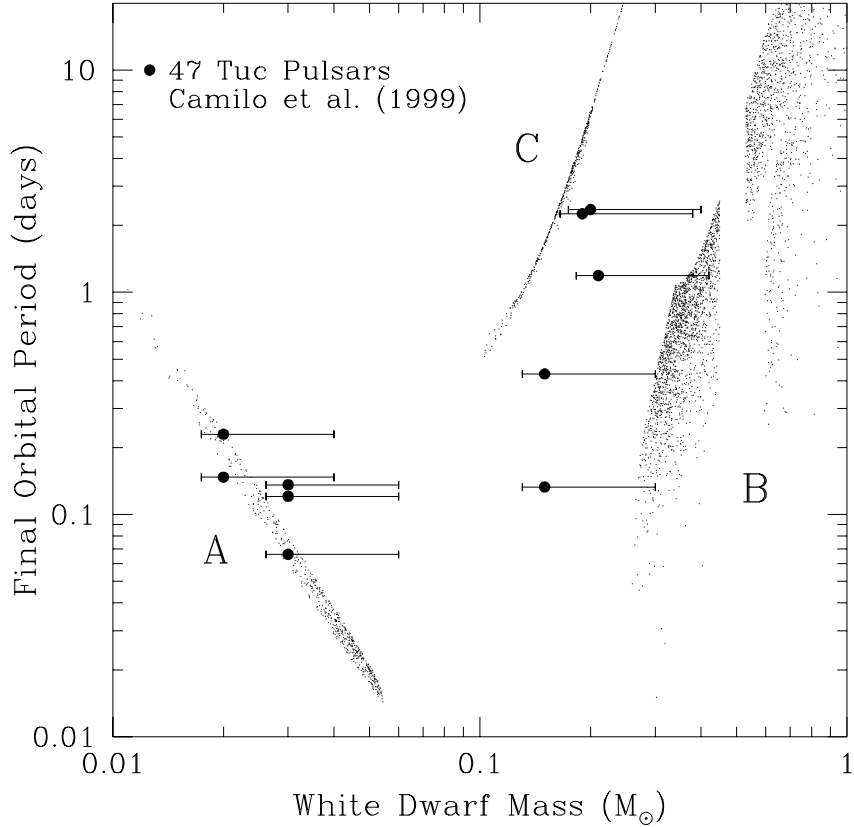}
\caption{Results of our initial Monte Carlo study of 
binary millisecond pulsar formation in a dense globular cluster such as 47~Tuc.  
Each small dot represents a binary
system in our simulation, while the filled circles are the 10 binary 
pulsars in 47~Tuc with
well measured orbits (the error bars extend from the minimum companion
mass to the 90\% probability level for random inclinations). 
There are 3 principal groups of simulated binaries.  
Systems in the
diagonal band on the left (A) are neutron star -- white dwarf binaries that decayed via
gravitational radiation to very short orbital periods ($\sim\,$mins), then
evolved with mass transfer back up to longer periods under the influence of
both gravitational radiation and tidal heating.  
The large group labeled B on the right
contains neutron star -- white dwarf binaries which had insufficient time to 
decay to Roche-lobe contact via the emission of gravitational radiation.  
The neutron stars in this group are not likely to be
recycled since they may not have accreted much
mass during the common envelope phase.
Finally, the systems lying
in the thin diagonal band toward longer periods (C) are those in which the mass
transfer from the giant or subgiant to the neutron star in the initial binary 
would be stable.  These have not
been evolved through the mass transfer phase; the mass plotted is simply
that of the He core of the donor star when mass transfer commences.  There
are many more systems in this category that have longer periods but lie off
the graph. See Rasio, Pfahl, \& Rappaport 2000 for more details.}
\end{figure}

We are also currently working on
incorporating into our Monte Carlo simulations
a more realistic treatment of tidal interactions, and, in particular,
tidal shocking through the Galactic disk (based on Gnedin, Lee, \& Ostriker 1999).
Tidal shocks can accelerate significantly both core collapse and the evaporation 
of globular clusters, reducing their lifetimes in the Galaxy (Gnedin \& Ostriker 1997).
Future work may include a fully dynamical treatment of all strong binary-single
and binary-binary interactions (exploiting the parallelism of the code to perform
separate numerical 3- or 4-body integrations for all dynamical interactions) as well
as a fully dynamical treatment of tidal shocking (performing short, $N$-body integrations
for each passage of the cluster through the Galactic disk or near the bulge).

\begin{acknowledgments}
This work was supported in part by NSF Grant AST-9618116 and NASA ATP 
Grant NAG5-8460.
Our computations were performed on the Cray/SGI Origin2000 
supercomputer at Boston University under NCSA Grant AST970022N.
\end{acknowledgments}

{

\bibliographystyle{apalike}

\begin{chapthebibliography}{1}

\bibitem{} 
Aarseth, S.J. 1999, PASP, 111, 1333

\bibitem{}
Bacon, R., Sigurdsson, S. \& Davies, M.B. 1996, MNRAS, 281, 830

\bibitem{}
Bailyn, C.D. 1995, ARA\&A, 33, 133

\bibitem{} 
Camilo, F., Lorimer, D.R., Freire, P., Lyne, A.G., \& Manchester, R.N. 2000,
  ApJ, 535, 975

\bibitem{} 
Chernoff, D.F. \& Weinberg, M.D. 1990, ApJ, 351, 121 

\bibitem{}
Cheung, P., Portegies Zwart, S., \& Rasio, F.A. 2001, in preparation

\bibitem{}
Djorgovski, S. 1993, in Structure and Dynamics of Globular Clusters,
eds. S.~Djorgovski \& G.~Meylan (ASP Conf.\ Series Vol.~50), 373

\bibitem{}
Edmonds, P.D., Grindlay, J.E., Cool, A., Cohn, H., Lugger, P., \& Bailyn, C. 1999,
 ApJ, 516, 250

\bibitem{} 
Fregeau, J.M., Joshi, K.J., Portegies Zwart, S., \& Rasio, F.A. 2001,
  in preparation

\bibitem{} 
Gao, B., Goodman, J., Cohn, H., \& Murphy, B. 1991, ApJ, 370, 567

\bibitem{} 
Gnedin, O.Y., \& Ostriker, J.P. 1997, ApJ, 474, 223

\bibitem{} 
Gnedin, O.Y., Lee, H.M., \& Ostriker, J.P. 1999, ApJ, 522, 935

\bibitem{}
Hansen, B.M.S., Kalogera, V., Pfahl, E., \& Rasio, F.A. 2001, ApJ, submitted

\bibitem{}
Heggie, D.C., Giersz, M., Spurzem, R., Takahashi, K. 1998, HiA, 11, 591;
for a more complete description of the experiment, go to \par
www.maths.ed.ac.uk/people/douglas/experiment.html

\bibitem{} 
H\'enon, M. 1971a, Ap.\ Space Sci., 13, 284

\bibitem{} 
H\'enon, M. 1971b, Ap.\ Space Sci., 14, 151

\bibitem{}
Hut, P., McMillan, S., Goodman, J., Mateo, M., Phinney, E.S.,
Pryor, C., Richer, H.B., Verbunt, F., \& Weinberg, M. 1992, PASP, 104, 981

\bibitem{} 
Joshi, K.J., Rasio, F.A., \& Portegies Zwart, S. 2000, ApJ, 540, 969 

\bibitem{} 
Joshi, K.J., Nave, C.P., \& Rasio, F.A. 2001a, ApJ, in press 
  [astro-ph/9912155]

\bibitem{} 
Joshi, K.J., Portegies Zwart, S., \& Rasio, F.A. 2001b, in preparation

\bibitem{}
Meylan, G., \& Heggie, D.C. 1997, Astron.\ Astrophys.\ Rev., 8, 1

\bibitem{} 
Portegies Zwart, S.F., \& McMillan, S.L.W. 2000, ApJ, 528, L17

\bibitem{} 
Rasio, F.A., Pfahl, E.D., \& Rappaport, S.A. 2000, ApJ, 532, L47

\bibitem{}
Sigurdsson, S., \& Phinney, E.S. 1995, ApJS, 99, 609

\bibitem{} 
Spitzer, L., Jr.  1969, ApJ, 158, L139

\bibitem{} 
Spitzer, L., Jr. 1987, Dynamical Evolution of Globular Clusters 
(Princeton: Princeton University Press) 

\bibitem{} 
Spitzer, L., Jr. \& Hart, M.H. 1971a, ApJ, 164, 399

\bibitem{} 
Spitzer, L., Jr. \& Hart, M.H. 1971b, ApJ, 166, 483

\bibitem{} 
Takahashi, K., \& Portegies Zwart, S.F. 1998, ApJ, 503, L49

\bibitem{}
Taylor, J.M., Grindlay, J.E., Edmonds, P.D., \& Cool, A.M. 2001, ApJ, submitted

\bibitem{} 
Watters, W.A., Joshi, K.J., \& Rasio, F.A. 2000, ApJ, 539, 331

\end{chapthebibliography}
\chapbblname{chapbib}
\chapbibliography{logic}

}

\end{document}